\documentstyle[aps,twocolumn,floats,prl]{revtex}
\input epsf

\newcommand{\el}{l}
\newcommand{\bn}{\hat{\bf n}}

\newcommand{\bsigma}{\mbox{\boldmath $\sigma$}}

\newcommand{\wj}[6]{\left(
                           \begin{array}{ccc}
        \! #1\! & #2\!  & #3\!  \\
        \! #4\! & #5\!  & #6\!
                           \end{array}
                   \right)}

\newcommand{\ApJL}{Astrophys. J Lett.}
\newcommand{\ApJ}{Astrophys. J}
\newcommand{\PRL}{Phys. Rev. Lett.}
\newcommand{\PRD}{Phys. Rev. D}
\newcommand{\MNRAS}{Mon. Not. Roy. Astr. Soc.}

\newcommand{\etal}{{\it et al.} }
\newcommand{\aut}[2]{{#2.\ #1,}}
\newcommand{\laut}[2]{{#2.\ #1,}}
\newcommand{\refs}[6]{#2, #3  {#4} (#5).}
\newcommand{\crefs}[6]{#2, #3  {#4} (#5);}
\newcommand{\urefs}[5]{#2, #3 (#4, #5).}
\newcommand{\curefs}[5]{#2, #3 (#4, #5);}
\newcommand{\mybib}[2]{\bibitem{#2}}

\begin{document}
\twocolumn[\hsize\textwidth\columnwidth\hsize\csname
@twocolumnfalse\endcsname

\title{Dark Synergy: Gravitational Lensing and the CMB}

\author{Wayne Hu}

\address{5640 S. Ellis Ave, University of Chicago, Chicago, IL 60637\\
Revised \today}
\maketitle
\begin{abstract}
Power spectra and cross-correlation measurements 
from the weak gravitational lensing of 
the cosmic microwave background (CMB) and the cosmic shearing of 
faint galaxies images
will help shed light on quantities hidden from 
the CMB temperature anisotropies: 
the dark energy, the end of the dark ages, 
and the inflationary gravitational wave amplitude.
Even with modest surveys, both types of lensing power spectra break CMB degeneracies 
and they can ultimately improve constraints on the dark energy equation of state $w$ by over an order of magnitude.
In its cross correlation with the integrated Sachs-Wolfe effect, CMB lensing offers
a unique opportunity for a more direct detection of the dark energy and enables study of
its clustering properties.  
By obtaining source redshifts and cross-correlations with CMB lensing, 
cosmic shear surveys provide tomographic 
handles on the evolution of clustering correspondingly better precision on the dark energy
equation of state and density.  Both can indirectly provide detections 
of the reionization optical depth and modest improvements in 
gravitational wave constraints which we compare to more direct constraints.
Conversely, polarization B-mode contamination
from CMB lensing, like any other residual foreground, 
darkens the prospects for ultra-high precision on gravitational
waves through CMB polarization requiring large areas of sky for statistical 
subtraction.
To evaluate these effects we provide fitting formula for the evolution 
and transfer function of the Newtonian gravitational potential.
\end{abstract}
\vskip 0.5truecm
]

\section{Introduction}

With the launch of the MAP satellite and continuing progress in
ground and balloon based experiments, 
cosmologists hope to soon be in a situation where cosmic microwave background
(CMB) anisotropies have firmly established the adiabatic cold dark matter
paradigm for structure formation and the parameters that govern it at high redshift.
Attention on the experimental and theoretical front will increasingly turn to the potentially 
deeper questions at the two opposite ends of time: the energy contents of the universe and their
clustering properties at recent epochs and the origins of structure 
perhaps in the inflationary epoch.   From distance measures to high redshift supernova
\cite{super} 
and indications of a near critical density universe from the CMB
\cite{flat}, there is increasingly
strong evidence for an unknown component of dark energy that accelerates the expansion at 
low redshifts.

In this context, it is useful to consider potential cosmological 
probes in light of
what the primary CMB temperature anisotropies are and are not
expected to reveal.  While some parameters such as the
physical baryon and non-relativistic matter density should be 
quite cleanly determined,
others such as the dark energy properties, the epoch of reionization, and the gravitational
wave amplitude are entangled with each other in parameter degeneracies.
While the CMB polarization is one well-recognized means of breaking some of these
degeneracies, these issues are sufficiently important and polarization measurements
sufficiently difficult that multiple independent approaches are desirable.

In this Paper, we compare and contrast the ability of weak gravitational lensing
in the shearing of faint galaxy images and distortions of the CMB temperature anisotropies 
in shedding light on these issues in the post-primary
CMB epoch. 
Weak gravitational
lensing shares with the primary anisotropies 
a unique status in cosmology in that its observables
are in principle predictable {\it ab initio} given a cosmological model. 
Measurements are limited mainly by instrumental systematics rather than 
unknown astrophysics.   As such lensing observables are well-suited to
complement information from the CMB.
 
Recent works \cite{ZalSel99,Zal00,Hu01} have shown that it is possible to map 
structures on the largest scales at high redshift through the lensing of the CMB.
We evaluate here the utility of such measurements and their cross-correlation with
the anisotropies themselves as well as cosmic shear for cosmological parameter estimation. 
On the cosmic shear side, we extend the work of \cite{HuTeg99} by considering
correlations with CMB temperature anisotropies and lensing.  
We also utilize the extended parameter space of \cite{Huetal99}
to study the background and clustering properties of the dark energy.
In this context, Huterer \cite{Hut01} has recently shown that the sub-arcminute
regime provides substantial information on the dark energy but will 
require a better understanding of the power spectrum and its statistical
properties in the deeply non-linear regime (e.g. \cite{WhiHu00}).  Here we 
take the complementary tack of supplementing
information in the translinear regime with source redshift information \cite{Hu99b}.

The outline of the paper is as follows: in \S \ref{sec:formalism} we describe the the cosmological parameter space, power spectra and cross correlations of the
observables and the Fisher formalism for parameter estimation forecasts.
In \S \ref{sec:phenomenology}, we discuss the phenomenology of the lensing observables
and their utility in breaking cosmological parameter degeneracies.  We present parameter
forecasts in \S \ref{sec:forecasts} and conclude in \S \ref{sec:discussion}.  In the
Appendix, we give fitting formula for the transfer function and evolution of the
Newtonian curvature in the presence of the dark energy.

\section{Formalism}
\label{sec:formalism}

We begin in \S \ref{sec:parameters} by defining the cosmological 
context paying special care to define quantities properly in the 
presence
of dark energy.  In \S \ref{sec:statistics} we review the harmonic 
formalism for handling scalar, vector and tensor fields on the sky
and consider the calculation of their power spectra and cross correlations
in \S \ref{sec:tracer}. Finally we generalize the Fisher formalism for
multiple observables from overlapping fields in \S \ref{sec:fisher}.

\subsection{Cosmological Parameters}
\label{sec:parameters}

We work in the context of spatially flat cold dark matter models for 
structure formation with initial curvature fluctuations. 
In units of the total (critical) density $3H^2/8\pi G$, with $c=1$,
the fractional contribution of each component is
denoted $\Omega_i(z)$, $i=c$ for the CDM, $b$ for the baryons and $\Lambda$
for the dark energy.  We also define the auxiliary quantity $\Omega_m = \Omega_c
+\Omega_b$, the total non-relativistic matter.  We assume throughout that the
neutrinos contribute negligible matter density.
The expansion rate at epochs where the radiation is negligible is given by
\begin{equation}
H^2 = H_0^2 [\Omega_m(0) (1+z)^3 + \Omega_{\Lambda}(0)
\rho_\Lambda(z)/\rho_\Lambda(0)] \,,
\end{equation}
where $H_0=100h$ km s$^{-1}$ Mpc$^{-1}$ is the Hubble constant.  The 
evolution
of the dark energy density is governed by its equation of state 
$w=p_{\Lambda}/\rho_{\Lambda}$ such that
\begin{equation}
\rho_\Lambda' = -3(1+w)\rho_\Lambda\,,
\end{equation}
where $'$ denotes a derivative with respect to $\ln (1+z)^{-1}$.  For 
illustrative purposes, we take $w=$ const. such that 
$\rho_\Lambda(z) = \rho_\Lambda(0) (1+z)^{3(1+w)}$.  Thus
4 parameters are associated with background energy densities 
$\Omega_b h^2$, $\Omega_m h^2$, $\Omega_\Lambda$
and $w$ all evaluated at the present epoch.   
We will often use the conformal lookback time in lieu of the redshift
\begin{equation}
D(z) = \int_0^z  {dz' \over H(z')} \,,
\end{equation}
abusing notation in the arguments of functions where no confusion will arise.
Overdots will represent derivatives with respect to $D$ throughout.
The final parameter associated with the background cosmology is the Thomson
optical depth in the reionization epoch $\tau$.

Four parameters are associated with the 
perturbations to the background.
An amplitude and a tilt define the initial fluctuations in 
the logarithmic power spectrum of
the Bardeen or comoving gauge curvature $\zeta$ 
\cite{Bar80}
\begin{equation}
\langle \zeta({\bf k},z) \zeta({\bf k}',z) \rangle = (2\pi)^3 \delta({\bf k} - {\bf k}') 
{2\pi^2 \over k^3} \Delta_\zeta^2(k,z)\,,
\end{equation}
as
\begin{equation}
\Delta_\zeta^2(k,z_i) = \delta_\zeta^2 \left( {k \over k_{\rm fid}} 
\right)^{n-1}\,,
\end{equation}
where the fiducial scale is taken to be $k_{\rm fid}=0.01$ Mpc$^{-1}$ and the 
initial (or inflationary) epoch $z_i$ is taken to be sufficiently 
early that all relevant scales are outside the horizon.  
Under slow roll inflation (e.g. \cite{LidLyt93}), 
\begin{equation}
\Delta_\zeta^2(k,z_i) = {8 \over 3 \epsilon} {V \over m_{\rm pl}^4} 
\,,
\end{equation}
where $m_{\rm pl}$ is the Planck mass and
$\epsilon = {3 \over 2}(1+w_i)$ is the deviation from vacuum domination in the
equation of state of the inflaton, all evaluated when the relevant 
scale exited the
horizon.

The Bardeen curvature provides a convenient representation since it 
remains constant outside the horizon in a flat universe regardless
of its energy contents.  It is related to
the power spectrum of Newtonian curvature fluctuations 
as
\begin{equation}
\Delta_\Phi^2(k,z) = \Phi_c^2(z) 
{T_w(k,z) \over T_w(0,z)} T_m(k) \Delta_\zeta^2(k,z_i)
\label{eqn:zetaphi}
\end{equation}
in the linear regime. 
The potential decay function in the large-scale limit $\Phi_c$ and transfer functions $T_w$ 
and $T_m$ are given in the Appendix.  The partitioning of the transfer function into
two pieces reflects its relationship
to the matter density fluctuations $\Delta_m^2$ in the comoving gauge 
\begin{equation}
\Delta_\Phi^2(k,z) = 
{9 \over 4} \left( {H_0 \over k} \right)^4 \Omega_m^2(0) (1+z)^2 \Delta_m^2(k,z)\,,
\end{equation}
where
\begin{equation}
\Delta_m^2(k,z) \approx \delta_H^2 \left( { k \over H_0} \right)^{3+n} \left( T_w(k,z) 
			{T_m(k) \over 1+z} {\Phi_s(z) \over \Phi_s(0)} \right)^{2}\,,
\end{equation}
under the approximation that the comoving dark energy density fluctuation contributes 
negligibly to the Newtonian curvature.
Here $\Phi_s$ is the potential decay function in the small scale limit
where dark energy clustering is negligible.
Note that in the presence of dark energy
with $w>-1$, the transfer function in the comoving gauge is no longer 
equivalent to that in the synchronous gauge on scales 
approaching $1/H(z)$ during dark energy domination.  The
factor $T_w(k,z)$ given in the Appendix accounts for dark energy clustering in the comoving gauge
(c.f. \cite{MaCalBodWan99}).
$T_m$ is the usual matter transfer 
function defined in the absence of dark energy clustering and is here numerically evaluated 
to $k \approx 0.5 h$ Mpc$^{-1}$ and 
extended to smaller scales with the fitting functions of 
\cite{EisHu99}.
We use the PD96 \cite{PeaDod96} scaling relations for $\Delta_m^2$ and the 
decay function $\Phi_s(D)$ to extend the potential power spectrum 
into the non-linear regime. 

These relations also 
give the mapping between our normalization scheme
and the more traditional one
\begin{equation}
\delta_H \approx {2 \over 3} \left( { k_{\rm fid} \over H_0 }\right)^{{1-n} \over 2} 
	{\Phi_s(0) \over \Omega_m(0) } \delta_\zeta\,.
\end{equation}

We allow for scale-invariant initial tensor or gravitational wave fluctuations.  
In the notation of
\cite{HuWhi97c}, where $H^{(\pm 2)} = (h_+ \mp i h_\times)/\sqrt{6}$ represent the amplitudes of the 
two polarizations in spin modes, the power spectrum in each component is given by
(e.g. \cite{LidLyt93})
\begin{eqnarray}
\Delta_H^2(k,z_i) &=& 
	\delta_T^2
\left( { k \over k_{\rm fid}} \right)^{n_T} 
 = {32 \over 9} 
		{V \over m_{\rm pl}^4}\,.
\end{eqnarray}
We take $n_T=0$ throughout.  

Finally, we take one parameter for the equation of state for the perturbations in the dark energy.
Because the dark energy has negative pressure, pressure fluctuations 
cannot be adiabatically related to density fluctuations through the 
equation of state.  Following \cite{Hu98}, we take
\begin{equation}
c_{\rm eff}^2 \equiv {\delta p \over \delta\rho} \Big|_{\rm rest}\,,
\end{equation}
to be the sound speed of the dark energy in its rest frame where its energy flux vanishes.
During dark energy domination the rest frame and the comoving frame 
coincide.
The dark energy can be considered smooth on scales smaller than the distance sound
can travel.  If the dark energy is composed of a single slowly rolling scalar field 
\cite{quintessence}, $c_{\rm eff}=1$
and the sound horizon coincides with the particle horizon.  Because the horizon at high 
redshift decreases, the clustering of the dark energy can leave an imprint on observable
scales even for a scalar field.  Measurement of this imprint can therefore test the scalar field 
paradigm for dark energy (see \S \ref{sec:cs}).
Note that as $w \rightarrow -1$, the phenomenological consequence of $c_{\rm eff}$ disappears
due to the vanishing of the relativistic energy flux $( \rho_\Lambda+p_\Lambda )v_\Lambda
\rightarrow 0$.  Dark energy candidates necessarily become indistinguishable from
a true cosmological constant in this limit.

This family of models is therefore described by 9 parameters.  We take as our
fiducial choices: $\Omega_b h^2 =0.02$, $\Omega_m h^2 = 0.148$, $\Omega_\Lambda=0.65$,
$w=-1$ or $-2/3$, $\tau=0.05$, $\delta_\zeta= 4.79 \times 10^{-5}$, $n=1$, $\delta_T =0$,
$c_{\rm eff}^2=1$.  
It is conventional to express the tensor amplitude in terms of
the scalar amplitude normalized to their individual contributions to the CMB temperature
quadrupole.  Note that the normalization is dependent on cosmological parameters, especially
the dark energy \cite{SchWhi01},  and we take the scaling appropriate to the $w=-1$ fiducial model:
\begin{eqnarray}
{T \over S}\Big|_{\rm fid} &=& \left( {\delta_T \over 1.85\times 10^{-5} 
}\right)^2\,,
\nonumber\\
			   &=& \left( {V^{1/4} \over 3.9 \times 10^{16} {\rm GeV}} \right)^4\,.
\end{eqnarray}
With this relation, $T/S$ constraints can be converted to tensor amplitude and inflationary
energy scale constraints.

\subsection{Observables and Power Spectra}
\label{sec:statistics}

CMB and lensing observables are described by scalar, vector and tensor 
fields on the sky.
A general scalar field on the sky $S(\bn)$, where $\bn$ is the directional vector,
is decomposed into multipole moments of the spherical harmonics
as
\begin{equation}
        S(\bn) = \sum_{\el m} S_{\el m} Y_{\el}^{m} (\bn)\,.
	\label{sec:scalarharmonics}
\end{equation}
Similarly a vector field ${\bf V}(\bn) = (V_1,V_2)$ is decomposed as
\begin{equation}
        [V_1 \pm i V_2](\bn) = \sum_{\el m} (P_{\el m} \pm i S_{\el m})\,
	{}_{\pm 1}Y_{\el}^{m} (\bn)\,,
	\label{eqn:vectorharmonics}
\end{equation}
where $S_{\el m}$ is the curl-free part and $P_{\el m}$ is the
divergence-free part. Here ${}_{s} Y_{\el m}$ are the spin-spherical 
harmonics \cite{spin}.
Finally a trace free tensor field can be represented
with the Pauli matrices ${\bf \bsigma}_{i}$ 
\begin{equation}
{\bf T}(\bn) = T_1 {\bf \bsigma}_1 + T_2 {\bf \bsigma}_2
+ T_3 {\bf \bsigma}_3\,;
\end{equation}
the trace behaves as a scalar field.
The symmetric part can be further decomposed as 
\begin{equation}
        [T_3 \pm i T_1](\bn) = \sum_{\el m}( S_{\el m}\pm i P_{\el m})\,
	{}_{\pm 2}Y_{\el}^{m} (\bn)\,.
	\label{eqn:tensorharmonics}
\end{equation}
For the tensor case $S_{\el m}$ is often called the ``electric'' or ``$E$'' and
$P_{\el m}$ the ``magnetic'' or ``$B$'' component of the field. 
The remaining piece can be decomposed as
\begin{equation}
	 T_2 (\bn) = \sum_{\el m} P_{\el m} Y_{\el}^m(\bn) \,,
\end{equation}
and is called the circular mode.
As the notation implies, the harmonics of the gradient and electric components
can be written in terms of those of a scalar potential field on the sky; 
the curl, magnetic and circular modes can be written in terms of 
the harmonics of a pseudo-scalar field. 

Statistical isotropy guarantees that the for any of two sets of 
harmonics $X = S,$ $P$
\begin{eqnarray}
    \left< X_{\el m}^{*} X'_{\el' m'} \right>
    &=& \delta_{\el,\el'}\delta_{m,m'} C_{\el}^{XX'} \,,
\label{eqn:powerdef}
\end{eqnarray}
which defines the power spectra.
Parity invariance requires that cross-spectra between scalar and 
pseudo scalar types
vanish.

\subsection{Tracer Fields}
\label{sec:tracer}

In the linear regime, all scalar fields on the sky that are related to cosmological structures 
can be thought of 
as line-of-sight projections of the gravitational potential $\Phi({\bf 
x},D)$ with a suitable weight 
\begin{equation}
X(\bn) = \int d D\, W^X(D) \Phi(D\bn,D) \,,
\label{eqn:scalardecomp}
\end{equation}
where 
$W$ can include differential operators on the potential field.  In the 
non-linear regime, any tracer of the density fluctuations may also be 
treated as such.  The scalar piece of vector and tensor fields can then 
also be reduced to this form.
 
Taking the harmonic moments of Eqn.~(\ref{eqn:scalardecomp}) yields,
\begin{eqnarray}
X_{\el m} &=& 4\pi i^\el \int {d^3 k \over (2\pi)^3} \Phi({\bf k},0) I_\el^X(k)
		Y_\el^m (\hat {\bf k})\,,
\nonumber\\
I_\el^X(k) &=& \int d D\, 
                { \Phi(k,D) \over \Phi(k,0) } W^{X}(k,D)
		j_\el (kD) \,.
\end{eqnarray}
The power spectrum of two fields then becomes
\begin{equation}
C_\el^{XX'} = 4\pi \int {dk \over k} I_\el^X(k) I_\el^{X'}(k) 
\Delta_\Phi^2(k,0)\,,
\label{eqn:clgen}
\end{equation}
and can be reexpressed in terms of the initial spectrum $\Delta_\zeta^2(k,z_i)$
through Eqn.~(\ref{eqn:zetaphi}).
For the CMB, this technique is 
known as the integral approach to anisotropies \cite{SelZal96}.

In the Limber approximation limit \cite{Kai92}, $k \gg \dot W^X/W^X$ and $\el \gg 1$,
\begin{equation}
I_\el^X(k) \approx 
\sqrt{\pi \over 2 \el} {1 \over k}
{ \Phi(k,\el/D) \over \Phi(k,0)} W^X(k,\el/k)\,,
\end{equation}
and with a change of variables $D= \el/k$ the power spectrum becomes
\begin{equation}
C_\el^{XX'} = {2\pi^2 \over \el^3} \int dD\, D W^X(D) W^{X'}(D) \Delta_\Phi^2(k,D)\,.
\end{equation}
We will use these equations to calculate the power spectra and cross 
correlations of the various effects.  

\begin{figure*}[htb]
\centerline{\epsfxsize=6truein\epsffile{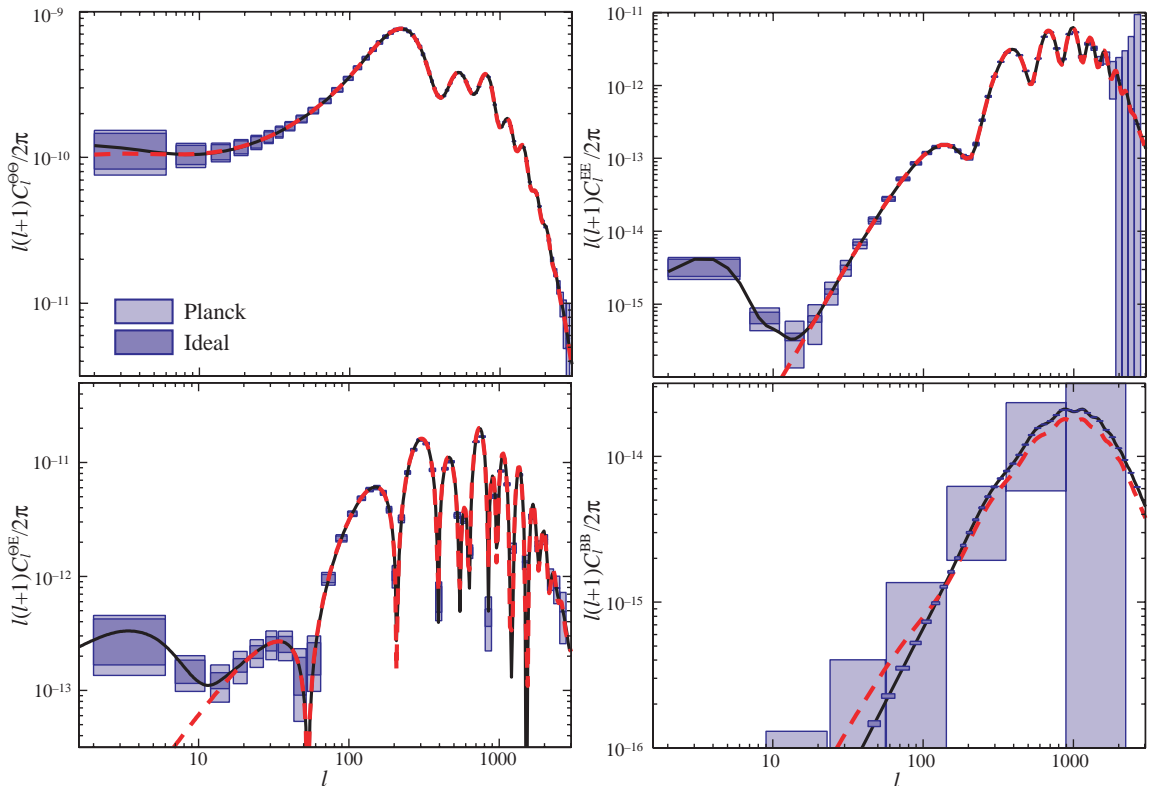}}
\caption{CMB power spectra in the fiducial model with $w=-1$ (solid) 
versus a dark energy model with $w=-2/3$ (dashed) and other parameters chosen 
to preserve the angular diameter distance and amplitude degeneracies 
(see text).  Boxes represent $1\sigma$ errors on band powers for the
Planck experiment and an ideal experiment out to $\el=3000$ (see 
Tab.~\protect\ref{tab:exp}). 
}
\vskip 0.25cm
\label{fig:cmb}
\end{figure*}

\subsection{Fisher Matrix}
\label{sec:fisher}

If all fields are Gaussian random, then the power and cross
spectra quantify all the information contained in the observables.
We can then use Fisher matrix techniques to combine, compare and 
contrast the statistical precision to which various surveys 
can determine the parameters underlying the power spectra.  

The Fisher matrix approximates the curvature of the likelihood function around
its maximum in a space spanned by the parameters ${\bf p}$  such
that the statistical errors on a given parameter $p_\alpha$: 
$\sigma(p_\alpha) \approx ({\bf F}^{-1})_{\alpha\alpha}$.
The usual formulae (e.g. \cite{TegTayHea97}) require a slight generalization to
account for the possibility that different surveys may only partially overlap in
sky coverage.  For the $i$th patch of sky, the elements of the
Fisher matrix are given by
\begin{equation}
F_{\alpha\beta}^i = \sum_{\el_{\rm min}}^{\el_{\rm max}}
		(\el + 1/2) f_{\rm sky}^i
 {\rm Tr}[ {\bf C}^{-1} {\bf C}_{,\alpha} 
                       {\bf C}^{-1} {\bf C}_{,\beta} ]\,,
\label{eqn:fisher}
\end{equation}
Here $,\alpha = \partial/\partial p_{\alpha}$ 
and ${\bf C}$ is the covariance matrix of the multipole moments
of the observables
\begin{equation}
{\bf C}_{X X'} = C_\el^{XX'} + N_\el^{XX'}
\end{equation}
where $N_\el^{XX'}$ is the power spectrum of the noise in the
measurement.  
$f_{\rm sky}^i$ is the
fraction of sky in the patch and quantifies the loss of independent modes
due to finite sky coverage.
We take $\el_{\rm min} = 0.5 f_{\rm sky}^{-1/2}$; the precise definition
does not matter due to the increase in sample variance on the scale of
the survey.  Although formally $\el_{\rm max} \rightarrow \infty$,
we generally take $\el_{\rm max}=3000$.  Above this scale non-Gaussianity
in both the CMB and lensing fields begin to violate the
assumptions behind the Fisher formalism.

Under the approximation that each patch is statistically independent,
the full Fisher matrix is the sum of those of the patches
\begin{equation}
F_{\alpha\beta} = \sum_{i=1}^{N_{\rm patch}} F_{\alpha\beta}^i\,.
\end{equation}

The parameters can consist of any set that suitably 
parameterizes the signal and noise power spectra. 
For example, they might be the signal power
spectra themselves in bands of $\el$.  
We use this parameterization when plotting the various 
observable power spectra  in \S \ref{sec:phenomenology}.  
They may alternately be the cosmological 
parameters described in \S \ref{sec:parameters}.  We take this 
approach
in \S \ref{sec:forecasts}.

\section{Phenomenology}
\label{sec:phenomenology}

Here we discuss the phenomenology of the various power spectra and cross correlations with
an emphasis on parameter degeneracies and dark energy.  
We begin with the CMB temperature field and proceed through CMB polarization, CMB lensing
and cosmic shear.  For each observable we give the statistical noise power spectra 
as functions of
experimental specifications. 

\subsection{CMB Temperature}
\label{sec:temp}

\begin{table}[tb]\footnotesize

\begin{center}
\begin{tabular}{rcccc}
Experiment & Chan. & FWHM & $\Delta T/T$ & $\Delta P/T$  \\
MAP
& 22 & $56'$ & 4.1 & 5.9 \\
$f_{\rm sky}=0.65$
& 30 & $41'$ & 5.7 & 8.0 \\
& 40 & $28'$ & 8.2 & 11.6 \\
& 60 & $21'$ & 11.0 & 15.6 \\
& 90 & $13'$ & 18.3 & 25.9 \\
Planck
& 30  & $33'$ & 1.6 & 2.3 \\
$f_{\rm sky}=0.65$
& 44  & $23'$ & 2.4 & 3.4 \\
& 70  & $14'$ & 3.6 & 5.1 \\
& 100 & $10.7'$ & 1.57 & 5.68 \\
& 143 & $8.0'$ & 2.0 & 3.7 \\
& 217 & $5.5'$ & 4.3 & 8.9 \\
& 353 & $5.0'$ & 14.4 & $\infty$ \\
& 545 & $5.0'$ & 147 & 208 \\
& 857 & $5.0'$ & 6670 & $\infty$ \\
$D_{4000}$
& 140 & $1.0'$ & 3.7 & $\infty$ \\
$f_{\rm sky}=0.1$ &&&&\\
Ideal
& --- & 0 & 0 & 0 \\
$f_{\rm sky}=1$ &&&&\\
\end{tabular}
\vskip 0.1truecm
\caption{CMB experimental specifications.  Channel frequency is given 
in GHz, FWHM in arcminutes and noise in $10^{-6}$.  The $D_{4000}$ is 
a mock up of a secondary CMB survey used for lensing
and the ideal experiment assumes
perfect information out to $\el=3000$.}
\label{tab:exp}
\end{center}
\end{table}

\subsubsection{Calculation}

The CMB temperature field 
$\Theta(\bn)=\Delta T/T$ 
is a scalar on the sky.  We calculate the CMB temperature power spectrum before lensing via
the Einstein-Boltzmann solver described in  \cite{Hu99b} based on the
hierarchy code of \cite{WhiSco96} and modified for dark energy.
Although the solutions may be recast into the integral form of 
Eqn.~(\ref{eqn:clgen}), the 
hierarchy technique provides better control over accuracy in the 
presence of degeneracies,
at the price of computational speed \cite{Huetal99}.
Gravitational lensing modifies the 
power spectrum \cite{Sel96,ZalSel98}, and we postprocess it following \cite{Hu00b}.   This power spectra is shown in Fig.~\ref{fig:cmb}
(top left).

\begin{figure}[htb]
\centerline{\epsfxsize=3.25truein\epsffile{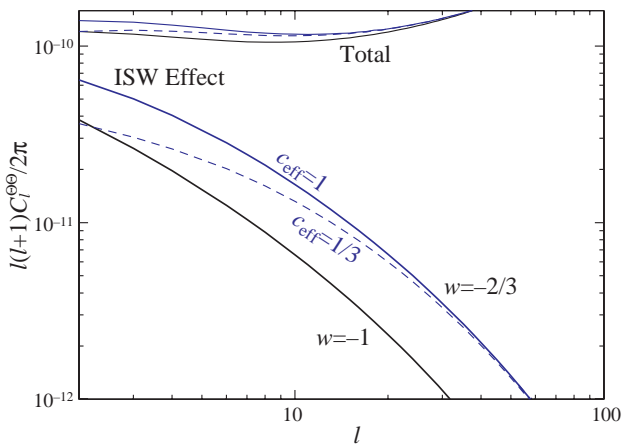}}
\vskip 0.25cm
\caption{ISW effect in the $w=-1$ fiducial model compared with models 
with
$w=-2/3$ and sound speeds $c_{\rm eff}=1,1/3$ with other parameters 
held fixed.   The ISW effect is highly sensitive to the equation of 
state
and clustering properties of the dark energy but only becomes a 
substantial
fraction of the total temperature anisotropy power spectrum at the lowest
multipoles.
}
\label{fig:isw}
\end{figure}

It will be useful to separate one contribution to the temperature 
anisotropies
for cross correlation studies.  
In the presence of dark energy, the decay of the Newtonian potential 
due to the inability of dark energy to cluster below its sound horizon 
produces a differential gravitational redshift whose net effect is 
called the Integrated Sachs-Wolfe (ISW) effect.  In a flat universe 
its
presence is a direct signature of dark energy.  Shown in 
Fig.~\ref{fig:isw} are the contributions as
calculated under the formalism of \S \ref{sec:tracer} with
\begin{equation}
W^{\Theta_{\rm ISW}}(D) = -2 {\dot \Phi \over \Phi} \,,
\end{equation}
and the growth rates given in the Appendix.   

Detector noise and telescope beam can be incorporated as a sky signal with
a 
spectrum given by the inverse variance weights of the channels
\begin{equation}
 (N_\el^{\Theta\Theta})^{-2} = \sum_{i=1}^{N_{\rm chan}}\left[
\left( {\Delta T \over T }\right)_{i} \sigma_i 
e^{\el(\el+1) \sigma_i/16\ln 2} 
\right]^{-2} \,,
\end{equation}
where $\sigma$ is the FWHM beam in radians. 
The noise and beam for various experiments are given in Tab.~\ref{tab:exp}.
In principle, foregrounds that are approximately Gaussian can also be
included in the noise term.  We will work in the idealization that they
are absent but see \cite{TegEisHudeO00} for potential effects of foregrounds
under the Fisher formalism.

\subsubsection{Degeneracies}

The Fisher matrix identifies degenerate directions in parameter
space through its eigenvectors.  It has been intensely studied
for primary CMB anisotropies \cite{forecasts,EisHuTeg99b} revealing 
two underlying and related degeneracies.
The first is the so-called angular diameter distance degeneracy.  
A change in parameters that leaves the angular diameter distance to the last
scattering surface at recombination {\it and} the physics of
acoustic oscillations unchanged preserves the structure and locations 
of the acoustic peaks.  In the present context, shifts to lower $\el$ by an
increase in $w$ can be compensated by a decrease in $\Omega_\Lambda$ as long
as the physical baryon and matter density are held fixed.  
The only way to break this degeneracy through the temperature spectrum is
to study the ISW contributions at the lowest $\el$'s. 

The large-scale nature of the ISW
effect is both a blessing and a curse. It offers the rare opportunity to study the 
properties of the dark energy including its clustering (see Fig.~\ref{fig:isw}). 
However precision in these studies is severely limited by sample
variance.  Even an all sky experiment has only a handful of realizations of the large
scale modes.  Worse still, as we shall see next, there are a multitude of effects
that can change the spectrum at the lowest $\el$'s.  
The angular diameter distance can be broken in two general ways:
with precision measures of a complementary combination of the parameters or through
isolation of the ISW effect.  The primary example of the former is external constraints
on the Hubble constant.  In the context of flat cosmologies, the CMB measurement of
$\Omega_m h^2$ combined with $h$-constraints yields a measure of $\Omega_\Lambda=1-\Omega_m$.

Because of the ISW effect, the angular diameter distance degeneracy is linked with
a degeneracy in the amplitude of the peaks relative to the lowest $\el$'s.  
The effect of reionization through $\tau$ is to uniformly lower the amplitude of the 
peaks compared with the lowest $\el$'s since scattering destroys anisotropies.  
It can therefore be 
compensated by a change in the initial amplitude $\delta_\zeta$ again except for 
the lowest $\el$'s.  Finally the tensor contribution also appears only at 
the lowest $\el$'s.  To resolve this degeneracy, the effects of reionization, 
initial amplitude, dark energy and tensors must be separated.  Of these
only reionization is likely to have direct external constraints, e.g. in
the form of a detection of the Gunn-Peterson effect.

In Fig.~{\ref{fig:cmb}} we show an example that employs both the angular diameter
distance degeneracy and the peak amplitude degeneracy.  The dashed line
represents a model with the parameters: 
$\Omega_b h^2=$ same, $\Omega_m h^2=$ same, $\Omega_\Lambda=0.54$, $w=-0.63$, $\tau=0$,
$\delta_\zeta=4.56\times 10^{-5}$, $T/S=0.015$.  From the unlensed temperature power spectrum
it is distinguished at only the  $0.2 \sigma$ level by the Planck experiment which is
essentially ideal for these purposes. 

\subsection{CMB Polarization}
\label{sec:pol}

The Stokes parameter
polarization fields for the linear polarization  of the
CMB  form a tensor field on the sky 
$T_1=U(\bn)$,
$T_2=V(\bn)=0$, and
$T_3=Q(\bn)$.
We define the corresponding multipole moments in 
Eqn.~(\ref{eqn:tensorharmonics})
as $E_{\el m}$ and $B_{\el m}$
for $E$ and $B$ modes respectively.
Their power spectra and cross-correlation with the temperature 
field are calculated in the same way as for the temperature 
anisotropies themselves. 
The effective noise power of an experiment is
given by 
\begin{eqnarray}
 (N_\el^{EE})^{-2} &=&  \sum_{i=1}^{N_{\rm chan}} \left[
\left( {\Delta P \over T }\right)_{i} \sigma_i 
e^{\el(\el+1) \sigma_i/16\ln 2} \right]^{-2} \nonumber\\
&=& (N_\el^{BB})^{-2}\,.
\label{eqn:pnoisepower}
\end{eqnarray}
We assume $N_\el^{\Theta E}=0$.
Values for various experiments are given in Tab.~\ref{tab:exp}.

As is well known, CMB polarization can break the peak amplitude degeneracy and so
also assist in breaking the angular diameter distance degeneracy.
Mainly, rescattering during reionization generates a  the low $\el$ bump in the
polarization $E$-power and $\Theta E$ cross spectra (see Fig.~\ref{fig:cmb}).  
Gravitational lensing and tensor fluctuations also generate $B$-mode polarization which
can help determine the initial amplitudes of the scalar and tensor fluctuations.
The main concern in this route to breaking parameter degeneracies is that the interesting
signatures are at the lowest $\el$'s where the polarization is at the level of {\it tenths} of a $\mu$K and below.  
The assumption
that foreground contamination is negligible compared with the sample 
errors on the fields
themselves is unlikely to hold true \cite{TegEisHudeO00}.  Note that in the context of constraints
on the tensor amplitude the gravitational lensing $B$-modes act as a foreground.
As we shall see in \S \ref{sec:tensors}, they place a lower limit on the detection threshold for tensors
even in the absence of true foregrounds.

\subsection{Lensing}
\label{sec:lensing}

The observables of weak lensing of the CMB and faint galaxies
are all based on the projected potential $\phi_{i}(\bn)$, a scalar field on the sky.
It follows the general prescription of a tracer field in \S \ref{sec:tracer}
with the lensing weight
\begin{equation}
    W^{\phi_{i}}(D)= {2 \over D}\int_{D}^{D(z_i)} dD'
                {(D'-D) \over D'} g_{i}(D')\,,
\label{eqn:lensingeff}
\end{equation}
from which one can calculate the multipole moments of $\phi_{i}$ 
and its cross-correlation
with other fields.  Here $g_{i}(D)$ is the source distribution for 
the $i$th 
set of lensed objects.

\begin{figure}[htb]
\centerline{\epsfxsize=3.25truein\epsffile{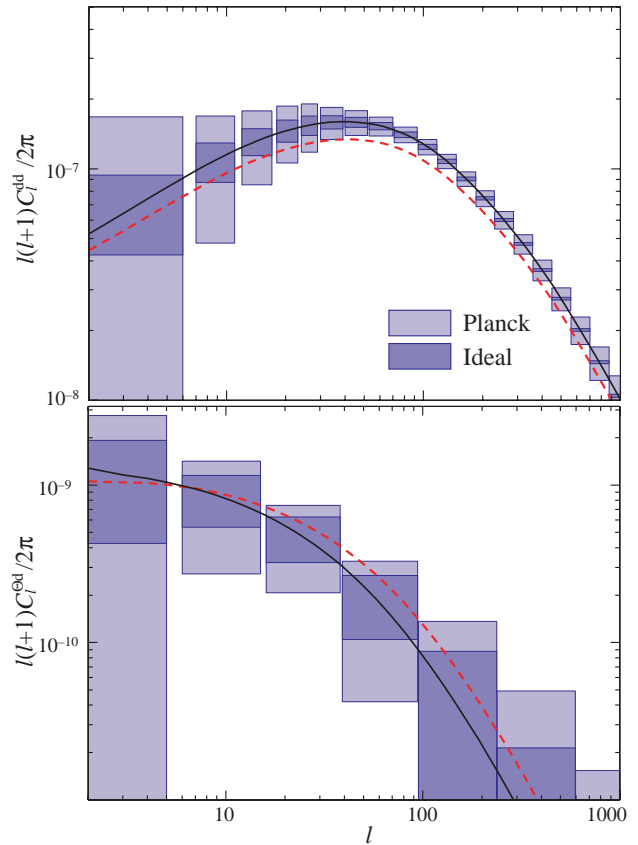}}
\vskip 0.25cm
\caption{CMB lensing power spectra for 
the fiducial $w=-1$ model (solid) and the degenerate
$w=-2/3$ model (dashed) of
Fig.~\protect\ref{fig:cmb}.  Boxes represent 1$\sigma$ errors on band
powers assuming the Planck and ideal experiments of 
Tab.~\protect\ref{tab:exp}.
Top: deflection power spectra. Bottom: cross correlation of deflection 
and temperature fields.
}
\label{fig:cmblens}
\end{figure}

\subsubsection{CMB Lensing}

For the CMB, it is the primary anisotropies themselves that are lensed
and the source distribution in Eqn.~(\ref{eqn:lensingeff}) is the Thomson 
visibility
\begin{equation}
g_{\rm CMB}(D) = \dot \tau e^{-\tau(D)}\,,
\end{equation}
where here and here only $\tau(D)$ refers to the optical depth out to
a distance $D$ and not the reionization optical depth.
It may be replaced by a delta function at the last scattering
surface $z \sim 10^3$.

The associated observable is 
the deflection angle
\begin{equation}
{\bf d}(\bn) = \nabla \phi_{\rm CMB}(\bn)\,,
\end{equation}
which remaps the original temperature field $\tilde\Theta$ as $\Theta(\bn) = 
\tilde\Theta(\bn+{\bf d})$ and similarly for the polarization field.
Its harmonic moments are thus curl free and obey
\begin{eqnarray}
[{d_1 \pm i d_2}](\bn) &=&\pm i \sum_{\el m} d_{\el m} 
	{}_{\pm 1} Y_\el^m (\bn) \,,\nonumber\\
   d_{\el m}  &=& -i \sqrt{\el (\el+1)} \phi_{\el m}\,.
\end{eqnarray}
These deflections alter the power spectrum of the temperature
and polarization fields.  On the scales of the acoustic peaks,
the main effect is a smoothing of features in the power spectra \cite{Sel96} and
the generation of $B$-mode polarization \cite{ZalSel98}.
These are potentially observable and can themselves break 
parameter degeneracies \cite{MetSil97,StoEfs99}.
One must be careful in that features can also be smoothed and $B$-modes generated
artificially by sky cuts and uneven sampling \cite{TegdeO01}.

The deflections also introduce non-Gaussianity into the CMB fields.
A negative impact of the non-Gaussianity is that creates a covariance between
the power spectra at different $\el$'s and technically invalidates the expression for the
Fisher matrix (\ref{eqn:fisher}).  The covariance is small and does not
affect the bulk of parameter estimation \cite{Hu01}.  However it can lead to misleadingly 
optimistic estimates of parameter forecasts when 
strong degeneracies like those discussed above are involved (see \S \ref{sec:forecasts}).

Fortunately, the non-Gaussianity also makes the deflection field itself 
and its power spectrum $C_\el^{dd}$ directly
observable with quadratic combinations of the temperature field. A 
quadratic estimator of the deflection field with the optimal noise 
power spectrum
\begin{eqnarray}
N_\el^{dd} &=&
	\left[ \sum_{\el_1 \el_2} {
        ( C_{\el_2}^{\tilde\Theta\tilde\Theta} F_{\el_1 \el \el_2} +
          C_{\el_1}^{\tilde\Theta\tilde\Theta} F_{\el_2 \el \el_1} )^2
          \over 2 (C_{\el_1}^{\Theta\Theta} + N_{\el_1}^{\Theta\Theta})
		  (C_{\el_2}^{\Theta\Theta} + N_{\el_2}^{\Theta\Theta}) }\right]^{-1}\nonumber\\
&&\times
	\el(\el+1) (2\el+1)\,,
	\label{eqn:deflectionnoise}
\end{eqnarray}
was given in \cite{Hu01} and involves the divergence of the 
temperature
weighted temperature-gradient field.  Here
$\tilde C_\el^{\Theta\Theta}$ is the unlensed CMB spectrum and
\begin{eqnarray}
F_{\el_1 \el \el_2}& =& \sqrt{ (2\el_1+1)(2\el+1)(2\el_2+1) \over 4\pi} 
\wj{\el_1}{\el}{\el_2}{0}{0}{0}
\nonumber\\
&& \times {1 \over 2} [\el(\el+1)+\el_2(\el_2+1)-\el_1(\el_1+1)]\,.
\end{eqnarray}
and is approximately Gaussian.
The deflection power spectrum for the fiducial model is shown in 
Fig.~\ref{fig:cmblens} (top) along with the degenerate model from 
Fig.~\ref{fig:cmb} and the band power errors calculated according to the noise 
spectrum of Eqn.~(\ref{eqn:deflectionnoise}).
Since the deflection strength depends on the absolute amplitude of 
the underlying potential, its power spectrum breaks the amplitude 
degeneracy of the CMB temperature fluctuations.  It also probes
the dark energy dependent growth rates and distances.

Because the deflections trace the gravitational potential, they are 
correlated with temperature anisotropies themselves through the ISW 
effect \cite{SelZal98i,GolSpe99,CooHu00a}.  The cross-power spectrum is shown in Fig.~\ref{fig:cmblens} 
(bottom).  It helps isolate the ISW contribution 
in the temperature anisotropies and provide a means of constraining
the clustering properties of the dark energy as we shall see in \S 
\ref{sec:cs}.

\begin{table}[tb]\footnotesize

\begin{center}
{\sc Lensing Survey Specifications}
\begin{tabular}{cccc}
Experiment & area & $N_{z}$ &
$\bar n_i$ \\
$W_{25}$    & 25	     & 1	& 56 \\
$Z_{25}$    & 25	     & 3	& (28,14,14) \\
$W_{1000}$  & 1000	     & 1	& 56\\
$Z_{1000}$  & 1000	     & 3	& (28,14,14) \\
$W_{65\%}$  & 27000	     & 1	& 56\\
$Z_{65\%}$  & 27000	     & 3	& (28,14,14) \\
\end{tabular}
\caption{Lensing survey specifications. Area is in deg$^{2}$, source
density in gal/arcmin$^{2}$ and 
median redshift $z=1$ corresponding to band divisions $z<1$, 
$1<z<1.5$ and $z>1.5$.}
\label{tab:wlexp}
\end{center}
\end{table}

\subsubsection{Cosmic Shear}
\begin{figure}[htb]
\centerline{\epsfxsize=3.25truein\epsffile{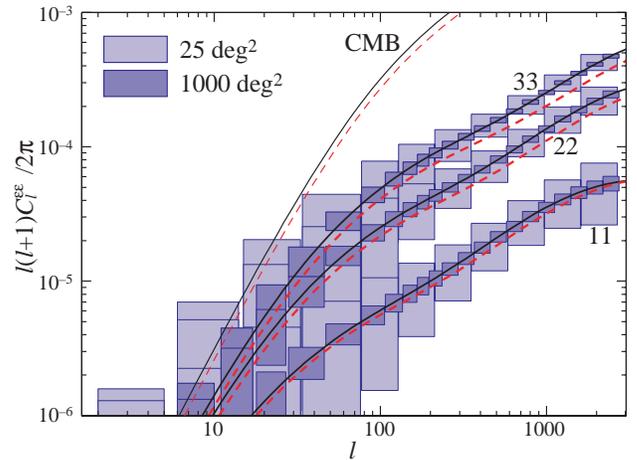}}
\vskip 0.25cm
\caption{Shear power spectra for three redshift bands $z<1$, $1< z < 
1.5$ and $z>1.5$ for the fiducial model (solid) and the degenerate 
$w=-2/3$ model of
Fig.~\ref{fig:cmb}.  Error boxes represent $1\sigma$ errors on band powers
appropriate to the survey parameters of 
Tab.~\ref{tab:wlexp}, $Z_{25}$ and $Z_{1000}$.  
}
\label{fig:w}
\end{figure}

\begin{figure}[htb]
\centerline{\epsfxsize=3.25truein\epsffile{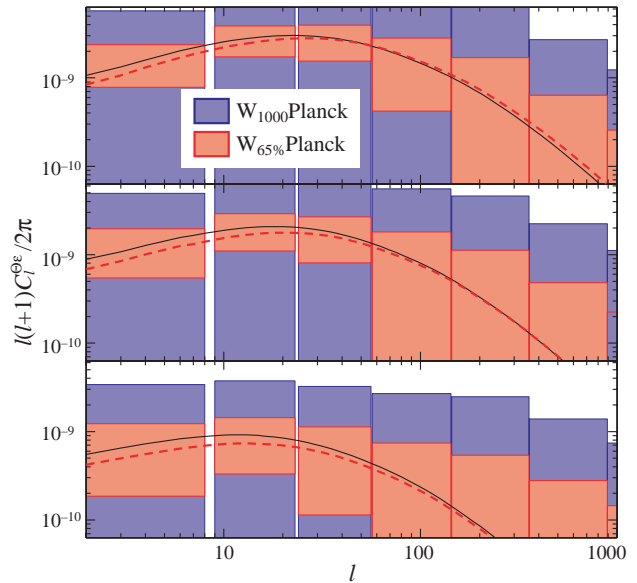}}
\vskip 0.25cm
\caption{Cross correlation of cosmic shear with the CMB temperature in 
three redshift bands for the fiducial model (solid) and the degenerate
$w=-2/3$ model of Fig.~\ref{fig:cmb}.
Errors are appropriate for Planck and lensing 
surveys with $1000$ deg$^{2}$ and all of the $65\%$ of sky covered by 
Planck. 
}
\label{fig:pxl}
\end{figure}

For galaxy weak lensing the
distance distribution of the sources is directly related to the source
galaxy redshift distribution,
\begin{equation}
g_{i}(D) = n_{i}(z) {d z \over dD}\,,
\end{equation}
where $n_{i}(z)$ is the normalized redshift distribution $\int dz  
n_{i}(z)=1$.
$n_{i}(z)$ is itself an observable that is produced in conjunction 
with the
survey but for definiteness we
take the redshift distribution corresponding to 
\begin{equation}
g_{\rm tot}(D) \propto D \exp[-(D/D_*)^4] \,,
\label{eqn:distribution}
\end{equation}
with $D_*$ fixed by the median redshift taken to be $z=1$. This 
distribution 
roughly approximates a survey with a magnitude limit of $R<25$.
For cosmic shear, the associated observable
is the symmetric trace free shear tensor
\begin{equation}
-[\nabla_i\nabla_j - {1 \over 2} g_{ij} \nabla^2]\phi(\bn)= [
\gamma_1(\bn)\sigma_3 + \gamma_2(\bn)\sigma_1]_{ij}\,,
\end{equation}
where $g_{ij}$ is the metric on the sphere.
Its harmonic moments are magnetic-mode free and obey
\begin{eqnarray}
[{\gamma_1\pm i \gamma_2}](\bn) &=&  \sum_{\el m} \epsilon_{\el m} 
	\,{}_{\pm 2} Y_\el^m (\bn) \,,\nonumber\\
   \epsilon_{\el m}  &=& -{1 \over 2} \sqrt{ (\el +2)! \over (\el -2)!}
	\phi_{\el m}\,.
\end{eqnarray}
Shot noise produces the noise power spectrum \cite{Kai92}
\begin{equation}
N_\el^{\epsilon\epsilon} = \langle \gamma_{\rm int}^2 \rangle / \bar n_{i}
\end{equation}
where $\langle \gamma_{\rm int}^2 \rangle^{1/2}$ is the rms intrinsic shear
per galaxy due to intrinsic ellipticities and measurement errors.  We
assume $\langle \gamma_{\rm int}^2 \rangle^{1/2} = 0.4$ throughout. $\bar n_i$
is the number of galaxies per steradian in the measurement.

With measurements of not just the redshift distribution but of 
individual source galaxies, the total 
distribution can be broken into redshift bands to yield separate but 
correlated power spectra.  The evolution of the spectra can be used
to probe structures and their evolution tomographically.  
To test the efficacy of tomography,
we divide the total into $N_{z}=3$ redshift bins that
contain a fixed fraction of the galaxies
(1) the lower half, (2) the third quartile and (3) the upper quartile
and label the distributions and as $g_{1}$, $g_{2}$ and $g_3$ respectively.
The shear power spectra and cross correlation in bands then follow from 
the prescriptions above.
This scheme was found in \cite{Hu99b} to be a good trade off
between shot noise and signal.  
Table~\ref{tab:wlexp} lists the parameters of the fiducial surveys used
in the Fisher analysis.

\begin{figure}[htb]
\centerline{\epsfxsize=3.25truein\epsffile{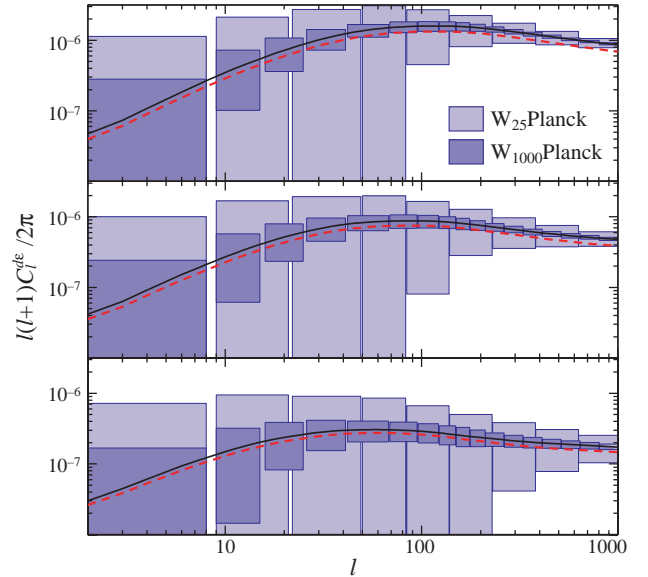}}
\vskip 0.25cm
\caption{Cross correlation of CMB deflection angle with cosmic shear in 
three redshift bands and errors appropriate for Planck and lensing 
surveys with $25$ and $1000$ deg$^{2}$.
}
\label{fig:dex}
\end{figure}

Similar to the CMB lensing case, the cosmic shear is correlated with 
the CMB temperature through the ISW effect as shown in 
Fig.~\ref{fig:pxl}.  Because the ISW effect is confined to 
low-$\el$'s, this correlation only becomes measurable with lensing 
surveys that cover a significant fraction of the sky. 
Finally the cosmic shear in the higher redshift bands 
and CMB deflection angles are substantially 
correlated as shown in Fig.~\ref{fig:dex}.  The CMB can thus provide
the high redshift anchor for tomography studies.

\begin{table*}[th]\footnotesize
\begin{center}
\begin{tabular}{lccccccccc}

&	
$\Omega_\Lambda$&	
$w$&	
$\tau$&	
$T/S$& 
$\ln \delta_\zeta$& 
$n$&	
$\ln\,\Omega_mh^2$&	
$\ln\,\Omega_bh^2$&	\cr
T&
    0.604&    1.93&    0.1833&    0.281&    0.1882&    0.0746&    0.1412&    0.0956\cr
TD&
    0.475&    1.42&    0.1684&    0.263&    0.1614&    0.0711&    0.1334&    0.0905\cr
TD$_{4000}$&
    0.195&    0.53&    0.0987&    0.142&    0.0549&    0.0474&    0.0719&    0.0621\cr
TP&
    0.330&    1.07&    0.0267&    0.139&    0.0448&    0.0360&    0.0696&    0.0493\cr
TPD$_{4000}$&
    0.162&    0.48&    0.0193&    0.084&    0.0208&    0.0185&    0.0160&    0.0300\cr
TH$_{10}$&
    0.083&    0.69&    0.1714&    0.262&    0.1832&    0.0721&    0.1369&    0.0937\cr
T$\tau_{10}$&
    0.565&    1.83&    0.0050&    0.280&    0.0834&    0.0733&    0.1401&    0.0947\cr
TW$_{25}$&
    0.295&    0.87&    0.1578&    0.110&    0.1122&    0.0450&    0.0745&    0.0577\cr
TZ$_{25}$&
    0.063&    0.29&    0.1297&    0.094&    0.0905&    0.0392&    0.0660&    0.0518\cr
TW$_{1000}$&
    0.083&    0.19&    0.0876&    0.077&    0.0658&    0.0230&    0.0435&    0.0326\cr
TZ$_{1000}$&
    0.010&    0.08&    0.0522&    0.065&    0.0426&    0.0125&    0.0288&    0.0222\cr
TPD$_{4000}$Z$_{1000}$&
    0.010&    0.04&    0.0141&    0.060&    0.0120&    0.0110&    0.0102&    0.0211\cr
\end{tabular}
\caption{Fisher parameter estimation errors for MAP and supplemented by 
various other sources. $T$ refers to temperature
spectra, $D$ deflection angles, $P$ polarization, $H$ 10\% Hubble constant measurements,
$\tau$ 10\% optical depth measurements ($\sigma(z_{i})\sim 0.5$),
$W$ weak lensing galaxy shear, $Z$ weak lensing galaxy shear with 3 redshift divisions.
Experimental assumptions are given in Tab.~\ref{tab:exp} and \ref{tab:wlexp}.  }
\label{tab:map}
\end{center}
\end{table*}

\begin{table*}[th]\footnotesize
\begin{center}
\begin{tabular}{lccccccccc}

&	
$\Omega_\Lambda$&	
$w$&	
$\tau$&	
$T/S$& 
$\ln \delta_\zeta$& 
$n$&	
$\ln\,\Omega_mh^2$&	
$\ln\,\Omega_bh^2$&	\cr
T&
    0.581&    1.88&    0.1724&    0.113&    0.1715&    0.0052&    0.0157&    0.0084\cr
TD&
    0.110&    0.35&    0.0262&    0.056&    0.0231&    0.0051&    0.0151&    0.0078\cr
TP&
    0.098&    0.32&    0.0042&    0.007&    0.0058&    0.0033&    0.0094&    0.0060\cr
TPD&
    0.065&    0.20&    0.0039&    0.007&    0.0054&    0.0030&    0.0079&    0.0056\cr
TH$_{10}$&
    0.070&    0.23&    0.1641&    0.106&    0.1635&    0.0051&    0.0157&    0.0082\cr
T$\tau_{10}$&
    0.553&    1.79&    0.0050&    0.086&    0.0086&    0.0051&    0.0156&    0.0082\cr
TW$_{25}$&
    0.265&    0.86&    0.0387&    0.057&    0.0340&    0.0051&    0.0152&    0.0079\cr
TZ$_{25}$&
    0.062&    0.20&    0.0313&    0.054&    0.0259&    0.0051&    0.0152&    0.0078\cr
TW$_{1000}$&
    0.050&    0.15&    0.0298&    0.053&    0.0240&    0.0050&    0.0148&    0.0078\cr
TZ$_{1000}$&
    0.010&    0.05&    0.0258&    0.053&    0.0208&    0.0046&    0.0135&    0.0076\cr
TPDZ$_{1000}$&
    0.010&    0.03&    0.0036&    0.007&    0.0039&    0.0026&    0.0026&    0.0045\cr
\end{tabular}
\caption{Same as Tab.~\ref{tab:map} but for Planck.}
\label{tab:planck}
\end{center}
\end{table*}

\begin{table*}[th]\footnotesize
\begin{center}
\begin{tabular}{lccccccccc}

&	
$\Omega_\Lambda$&	
$w$&	
$\tau$&	
$T/S$& 
$\ln \delta_\zeta$& 
$n$&	
$\ln\,\Omega_mh^2$&	
$\ln\,\Omega_bh^2$&	\cr
T&
    0.451&    1.45&    0.1343&    0.090&    0.1335&    0.0017&    0.0020&    0.0012\cr
TD&
    0.050&    0.16&    0.0077&    0.041&    0.0079&    0.0016&    0.0020&    0.0011\cr
TP&
    0.049&    0.16&    0.0015&    0.000&    0.0018&    0.0009&    0.0008&    0.0004\cr
TPD&
    0.018&    0.06&    0.0015&    0.000&    0.0017&    0.0009&    0.0008&    0.0004\cr
TH$_{10}$&
    0.069&    0.22&    0.1299&    0.084&    0.1292&    0.0016&    0.0020&    0.0011\cr
T$\tau_{10}$&
    0.435&    1.40&    0.0050&    0.065&    0.0054&    0.0016&    0.0020&    0.0011\cr
TW$_{25}$&
    0.248&    0.80&    0.0248&    0.045&    0.0245&    0.0016&    0.0020&    0.0011\cr
TZ$_{25}$&
    0.062&    0.20&    0.0129&    0.042&    0.0130&    0.0016&    0.0020&    0.0011\cr
TW$_{1000}$&
    0.047&    0.15&    0.0059&    0.041&    0.0059&    0.0016&    0.0020&    0.0011\cr
TZ$_{1000}$&
    0.010&    0.03&    0.0043&    0.041&    0.0046&    0.0016&    0.0020&    0.0011\cr
TPDZ$_{1000}$&
    0.008&    0.03&    0.0012&    0.000&    0.0015&    0.0009&    0.0007&    0.0004\cr
\end{tabular}
\caption{Same as for Tab.~\ref{tab:map} but for an ideal CMB experiment out to $\el=3000$.}
\label{tab:ideal}
\end{center}
\end{table*}
\section{Parameter Forecasts}
\label{sec:forecasts}

Here we study parameter forecasts using the Fisher matrix formalism of
\S \ref{sec:fisher} to combine information
from the primary CMB anisotropies and gravitational lensing.   We give details of the implementation in 
\S \ref{sec:methodology} and discuss the effect of lensing on the gravitational wave and
reionization detectability in \S \ref{sec:tensors} and dark energy properties in 
\S \ref{sec:eos}-\ref{sec:cs}.

\begin{figure}[htb]
\centerline{\epsfxsize=3.25truein\epsffile{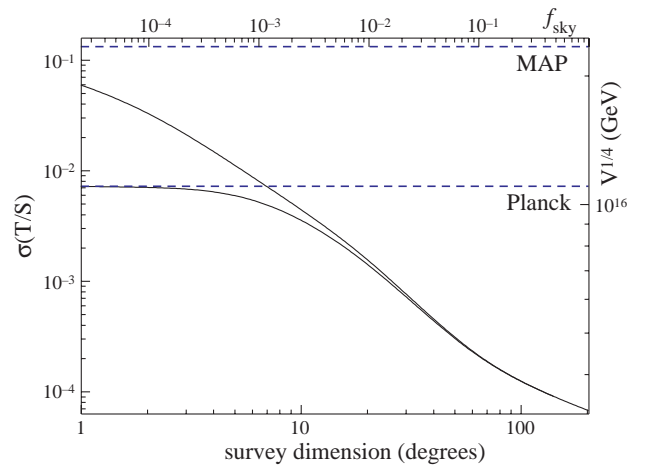}}
\vskip 0.25cm
\caption{Improvement in the polarized MAP and Planck $1\sigma$ detection 
thresholds for tensors with a dedicated polarization survey.
The statistical subtraction of
the lensing $B$-mode contamination requires a large survey area and places
and ultimate detection threshold of $3-4\times 10^{15}$ GeV for the
energy scale of inflation.
}
\label{fig:bimprove}
\end{figure}

\subsection{Methodology}
\label{sec:methodology}

The methodology of Fisher-matrix parameter forecasts with the CMB and cosmic shear
are well established \cite{forecasts,EisHuTeg99b,Hu99b}.  Here we simply note 
the details of our implementation.  We approximate the parameter derivatives
in the Fisher matrix (\ref{eqn:fisher}) with finite differences of step size
$\Delta \Omega_b h^2 = \pm 0.15\Omega_b h^2$, $\Delta \Omega_m h^2 = \pm 0.05 \Omega_m h^2$,
$\Delta \Omega_\Lambda = \pm 0.05 \Omega_m$, $\Delta n= \pm 0.005 n$, $\Delta \delta_\zeta 
=\pm 0.1 \delta_\zeta$, $\Delta \tau = \pm 0.1 \tau$, $\Delta w = 0.1w$, $\Delta \log_{10} c_s = -2$,
where ``$\pm$'' refers to the fact that two-sided differences are taken for better accuracy.
Derivatives with respect to $\delta_T$ or $T/S$ are simply proportional to the power spectra
themselves since non-linearities never develop in the tensor sector.  For the fiducial
model of $w=-1$, derivatives with respect to the sound speed vanish identically and consequently
these elements are dropped from the Fisher matrix.
We truncate the Fisher sum in Eqn.~(\ref{eqn:fisher}) at $\el_{\rm max}=3000$;
beyond this secondary anisotropies and nonlinearities in the projected potential
make the associated CMB and lensing observables non-Gaussian and invalidate
the Fisher formalism.

\begin{figure}[htb]
\centerline{
\epsfxsize=3.25truein\epsffile{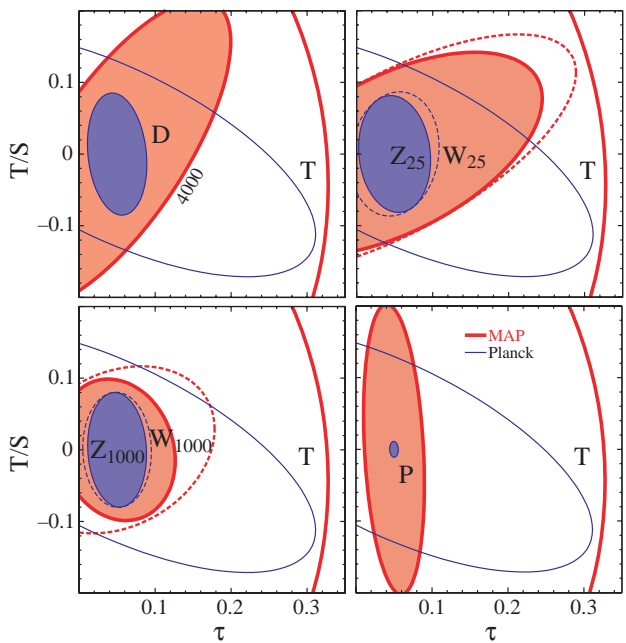}}
\vskip 0.25cm\caption{
Breaking of the tensor-reionization degeneracy.  Top-left: addition of CMB
deflection angle information (``D'') to the MAP (thick) and Planck (thin) temperature
constraints.  For MAP, we assume that the deflection angle information comes
from a separate $4000$deg$^{2}$ secondary anisotropy survey.  For Planck,
we assume that they are internal. Top-right: the addition of a $25$ deg$^2$
cosmic shear survey with ($Z_{25}$) and without ($W_{25}$) tomographic redshift
information.  Bottom-left: same but for a $1000$ deg$^2$ cosmic shear survey.
Bottom-right: addition of polarization information.
}
\label{fig:ts}
\end{figure}

As discussed in \cite{EisHuTeg99b}, the angular diameter distance degeneracy must be
protected against numerical errors.  We replace finite differences in $w$ with those
in $\Omega_\Lambda$ beyond $\el=150$ with the proportionality fixed at this scale.  
We have tested that the results are insensitive to the exact choice of the matching.

For the CMB power spectra, we have the choice of using the lensed or unlensed power spectra
as inputs to the Fisher matrix.  As discussed in \S \ref{sec:phenomenology}, using the
unlensed spectra generally underestimates the information content since lensing breaks parameter
degeneracies whereas using the lensed power spectra overestimates the information content due
to the non-Gaussian correlation of power spectra errors.  One can show that using the lensed
power spectra for a ideal experiment out to $\el=3000$ and Gaussian assumptions artificially
predicts a better breaking of the angular diameter distance degeneracy than 
complete (cosmic variance limited) information on both the unlensed power 
spectrum and the deflection angles.  The reason is
that lensing effects at $\el \sim 1000$ still arise from mass structures
at $\el \sim 100$.  Consequently the sample variance on lensing effects
is much larger than the Gaussian assumption would imply.  For this reason and
the fact that we directly measure the deflection spectrum through quadratic statistics, we
use the unlensed power spectra in parameter forecasts given in Tab.~\ref{tab:map}-\ref{tab:ideal}.  
The exception is in the discussion of
the tensor amplitude and $B$-mode polarization in \S \ref{sec:tensors}.  Here the generation of
$B$-modes by lensing introduces a foreground to the tensor measurement and the unlensed spectra
would give a falsely optimistic limit on the detectability of tensors.

\subsection{Tensors and Reionization}
\label{sec:tensors}

Gravitational lensing both provides and obscures information about the 
tensor or gravitational wave fluctuations. 
In the absence of lensing and with the complete removal of foregrounds through their
frequency dependence, the $B$-mode of the CMB polarization maps provide a direct
measure of the tensor contribution that is ultimately limited only by its own cosmic
variance. 
By generating $B$-modes in the polarization
with a blackbody spectrum, lensing adds an extra source of noise bias that must be
subtracted statistically.  Hence the threshold for tensor detectability is set by
the sample variance of the lensing not the tensor signal.   Without lensing 
(or foregrounds and systematics) it is always better to go deep on a small
patch than shallow on a wide patch.  The optimal size is approximately
$3^\circ \times 3^\circ$ \cite{JafKamWan99}. 
With lensing, more samples of such regions is required to beat down the
variance on the lensing contamination if extremely small tensor signals
are to be recovered.

To quantify these considerations, we use the Fisher approach to examine the
$1\sigma$ threshold for detection of tensors including the lensed polarization
as a Gaussian random field.
For the detector noise limited, all-sky MAP and Planck missions lensing has
essentially no effect on the detectability of tensors.  

Lensing does change the optimal strategy for 
a dedicated polarization experiment that seeks to improve on the Planck
experiment as shown in Fig.~\ref{fig:bimprove}.  
To reach below $T/S\approx 0.01$ (or inflationary energy scales $< 10^{16}$ 
GeV)
and improve on Planck's potential, a survey area
of greater than $10^\circ \times 10^\circ$ is required.  

Note that these considerations assume perfect foreground and systematic error
removal (including $E$, $B$-mode separation in a finite survey) as well as
a Gaussian $B$-field from lensing.  As such, they should be taken as a lower limit
on the detectability of tensors.  Conversely, they assume statistical
subtraction only; limits can be improved if direct subtraction methods
can be developed.

Lensing also indirectly assists the detection of tensors in the absence
of polarization.  Since lensing is sensitive to the absolute amplitude of the potential
fluctuations, measurements of the CMB deflection power spectrum or cosmic shear
power spectra can break the amplitude degeneracy of the CMB acoustic peaks and
so improve the errors on both tensors and the reionization optical depth.

\begin{figure*}[htb]
\centerline{
\epsfxsize=5truein\epsffile{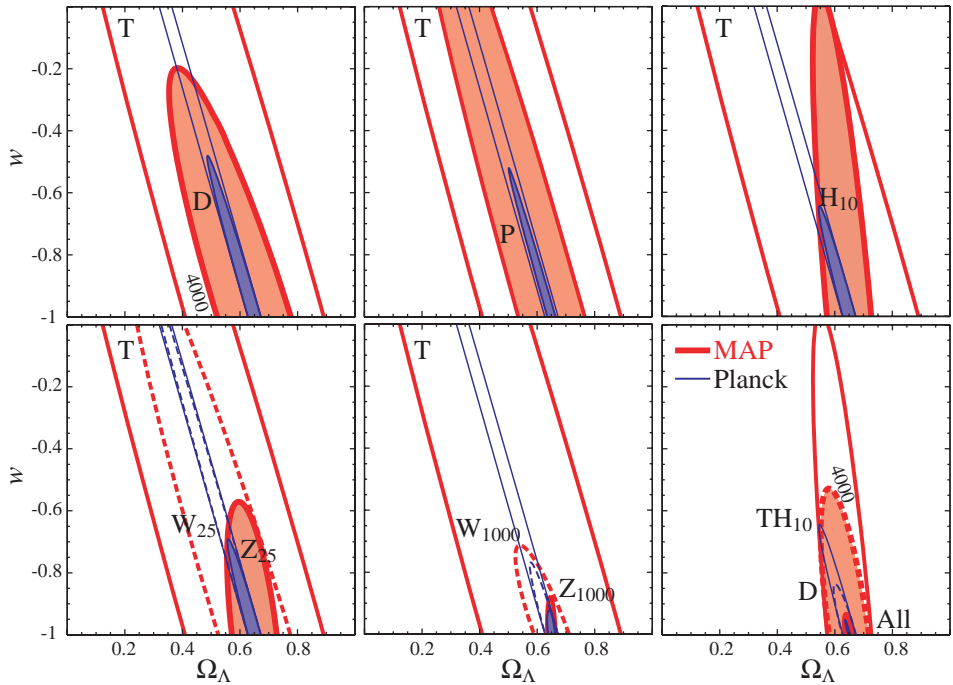}
	}
\vskip 0.25cm
\caption{Improvement on the MAP (thick) and Planck (thin) temperature ($T$)
determination of the dark energy
equation of state and density.  Clockwise from the top left: 
addition of CMB deflection angles ($D$);
polarization ($P$); 10\% Hubble constant measurements ($H$); 25 deg$^2$ cosmic shear survey 
with ($Z$ solid) and without ($W$ dashed) tomography; same but for 1000 deg$^2$; Hubble constant ($TH$),
plus deflections ($D$), plus 
a 1000 deg$^2$ lensing survey (All).
}
\label{fig:ellipsew}
\end{figure*}

In Fig.~\ref{fig:ts}, we quantify this degeneracy breaking.  While polarization information
still provides better constraints on tensors and reionization, deflection angle information
can improve errors on $\tau$ by $2-20$ (MAP to Ideal CMB experiment) 
and $T/S$ by $2$. 
Cosmic shear can help by a 
comparable but somewhat smaller amount with or without tomographic 
information.  The reionization epoch is also potentially directly observable
in the Gunn-Peterson effect and so we show the influence of a prior of
$\sigma(\tau)=0.1\tau$ on the other parameters in 
Tab.~\ref{tab:map}-\ref{tab:ideal}.

\subsection{Equation of State}
\label{sec:eos}

As is well known and shown in Fig.~\ref{fig:ellipsew}, there is an angular diameter 
distance degeneracy between the dark energy equation of state $w$ 
and energy density $\Omega_\Lambda$.  
There are many ways to break the angular diameter distance degeneracy some involving
pure geometry and other employing the clustering properties of the dark matter
and dark energy.  As such strong consistency checks will be available
for parameter constraints and underlying assumptions for dark energy parameters.

Although both MAP and Planck show a strong
degeneracy, it is important to note that for the Planck experiment the direction orthogonal
to the degeneracy line is highly constrained.  This corresponds to the better
constraints
on $\Omega_m h^2$ which also enters into the angular diameter distance relation. 

A purely geometric way of breaking the degeneracy then is to introduce constraints on 
the Hubble constant.  In a flat universe, a precise determination of $\Omega_m h^2$ combined
with constraints on $h$ yield corresponding constraints on $\Omega_\Lambda=1-\Omega_m$ as shown
in Fig.~\ref{fig:ellipsew}.  For the Planck experiment, the 10\% measurement of the Hubble
constant currently claimed \cite{Freetal01} is sufficient to yield an interesting constraint
on the equation of state $\sigma(w)=0.23$ (see Tab.~\ref{tab:planck}).
Using the baryon bumps in the galaxy power spectrum as a standard ruler to measure the
Hubble constant, this means of degeneracy breaking can potentially be substantially
improved \cite{EisHuTeg99b}.

As seen in Fig.~\ref{fig:cmblens}, the CMB deflection power spectrum is another means
of breaking the degeneracy.  It differs by also involving the effect of the dark energy on
the clustering of the matter.  Because of the nature of the quadratic estimator
of the deflection angle, it is crucial here to resolve CMB temperature anisotropies
through the damping tail to $\el \sim 3000$ \cite{Hu01}.  This is reflected in the
negligible improvement in $\sigma(w)$ for MAP alone to the order of magnitude improvement 
for the ideal experiment.

Information on the deflection power spectra do not
have to come from the same experiment as that for the temperature anisotropies themselves.
To measure deflection angles, one requires high resolution in the temperature map but
essentially no information on the large-scale anisotropy itself.  Combining an all sky
experiment such as MAP with
an experiment that is dedicated to measuring secondary arcminute scale anisotropies
can therefore be fruitful.  We show in Fig.~\ref{fig:ellipsew}
that a $4000$ deg$^2$ survey is sufficient to provide interesting constraints on the
equation of state.

Similarly cosmic shear power spectra also provide information on the equation of state.
As shown by \cite{Hut01}, if the whole power spectrum can be recovered 
to $\el=10000$ and theoretical predictions in the deeply non-linear
regime improved, a single redshift band suffices to yield powerful constraints on
the equation of state in the Gaussian approximation.   Non-linearities produce
non-Gaussianity in the cosmic shear that degrades the amount of information in the
deeply non-linear regime beyond $\el \sim 3000$ \cite{CooHu01}.  In Fig.~\ref{fig:ellipsew}
we show that information in the trans-linear regime of $\el < 3000$ suffices to determine
the dark energy equation of state when broken into multiple redshift bands and combined
with CMB temperature information.  Notice that redshift information on a 25 deg$^2$ survey
(filled ellipses) is competitive with no redshift information on a 
1000 deg$^2$ survey (dashed ellipses).

With the multitude of avenues for constraints on the dark energy equation of state
discussed above as well as those from high redshift supernovae \cite{super}
and number counts \cite{HaiMohHol01},
it is possible that the observations will be inconsistent with the simple underlying
model of a constant equation of state and dark energy clustering appropriate
for a single
slowly-rolling scalar field.  Since geometric tests can potentially probe the
time evolution of the equation of state, we conclude in the next section with a discussion of
dark energy clustering.

\begin{figure}[htb]
\centerline{\epsfxsize=3.25truein\epsffile{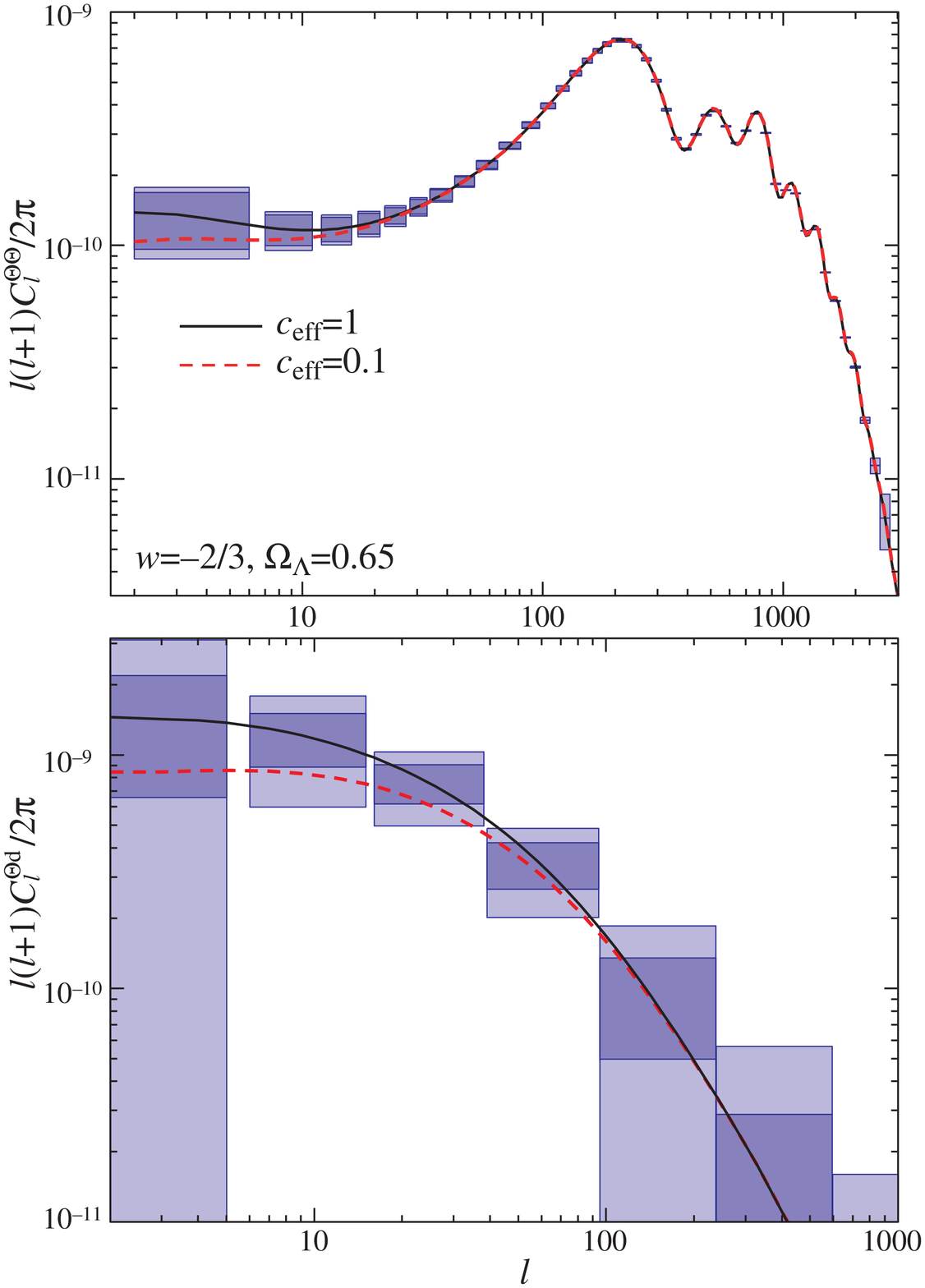}}
\vskip 0.25cm
\caption{Dark energy clustering in a model with $w=-2/3$ and $c_s=1$, $0.1$ 
and other parameters the same as in the fiducial model.  Top: effect on the 
CMB. 
Bottom: cross correlation of deflection angles with the temperature
anisotropies.
}
\label{fig:cs}
\end{figure}

\begin{table*}[th]\footnotesize
\begin{center}
\begin{tabular}{lcccccc}
& \multicolumn{3}{c}{Planck}
& \multicolumn{3}{c}{Ideal} \cr
&	
$\Omega_\Lambda$&	
$w$&	
$\log_{10}c_{\rm eff}$&
$\Omega_\Lambda$&	
$w$&	
$\log_{10}c_{\rm eff}$
\cr
T&
    0.4377&    1.105&    4.1221&
    0.3527&    0.890&    3.2039\cr
TD&
    0.0861&    0.215&    1.0938&
    0.0329&    0.083&    0.4973\cr
TP&
    0.1967&    0.497&    1.1856&
    0.1085&    0.274&    0.8531\cr
TPD&
    0.0431&    0.099&    0.7516&
    0.0088&    0.021&    0.3828\cr
TH$_{10}$&
    0.0693&    0.179&    3.8768&
    0.0687&    0.173&    3.0108\cr
TW$_{25}$&
    0.2377&    0.599&    1.4426&
    0.2201&    0.556&    1.1640\cr
TZ$_{25}$&
    0.0623&    0.158&    1.3827&
    0.0618&    0.156&    1.0978\cr
TW$_{1000}$&
    0.0479&    0.114&    1.3599&
    0.0447&    0.113&    1.0798\cr
TZ$_{1000}$&
    0.0106&    0.030&    1.3497&
    0.0099&    0.025&    1.0724\cr
TPDZ$_{1000}$&
    0.0098&    0.023&    0.7264&
    0.0058&    0.014&    0.3790\cr
\end{tabular}
\caption{Fisher parameter estimation errors for dark energy parameters
in a fiducial $w=-2/3$ model. Notation follows Tab.~\ref{tab:map}. Parameters
not shown are marginalized.} 
\label{tab:cs}
\end{center}
\end{table*}
\subsection{Dark Energy Clustering}
\label{sec:cs}

If the equation of state of the dark energy $w>-1$, then there is a new dimension to the
dark energy defined by its clustering properties.  In \S \ref{sec:parameters}, we introduced
the sound speed of the dark energy for this purpose.  Recall that the scalar field candidate
for the dark energy has $c_{\rm eff}=1$.  

As shown in Fig.~\ref{fig:isw} and \ref{fig:cs}, the ISW effect in the CMB rapidly decreases 
with the sound speed but is difficult to isolate from other contributions to the anisotropies
at low $\el$.  By breaking the amplitude degeneracy, the deflection power and
cosmic shear power spectra help isolate the ISW effect.
Furthermore the deflection angles
are themselves correlated with the temperature anisotropies leading to an additional
more direct handle on the dark energy clustering (see Fig.~\ref{fig:cs} bottom).  For
Planck the constrains are are equivalent to saying the dark energy is smooth at least across
$\sim$10\% of the current horizon or 1.4 Gpc in the fiducial model.  
As Tab.~\ref{tab:cs} shows, there
is room for substantial improvement especially on the CMB deflection angle side for a next
generation mission with higher angular resolution.  With an ideal CMB experiment to $\el=3000$
and a cosmic shear survey with $1000$deg$^2$, the dark energy smoothness 
an be constrained to be $\sim$40\% of the current horizon or 6 Gpc.   If cosmic shear surveys
can reach the sky coverage and control of systematics to measure the multipoles $\el \ll 100$
then additional information and consistency checks will be available
from their cross-correlation with the CMB temperature
maps (see Fig.~\ref{fig:pxl}).  Note however that these constraints greatly weaken as the equation of state approaches
$-1$.

\section{Discussion}
\label{sec:discussion}

Gravitational lensing as manifest in CMB deflection and cosmic shear
measurements
complements CMB primary anisotropies by providing information that
breaks degeneracies involving the dark energy density and equation of state, reionization 
and gravitational waves, specifically the angular diameter distance degeneracy and the 
amplitude degeneracies in the acoustic peaks.  In this way, it is similar in utility to the
well-studied CMB polarization and offers sharp consistency checks on the difficult-to-measure
dark energy parameters.  Conversely, CMB lensing obscures polarization
information on the gravitational waves and necessitates large sky coverage to
beat down sample variance even with perfect detectors and no foregrounds.

CMB lensing offers information that is similar to cosmic shear but with important
additional strengths and weaknesses.   Its primary strengths are that it is intrinsically
more sensitive to structure on larger scales and higher redshifts than even the 
next generation of wide-field galaxy surveys.  
These strengths translate into the opportunity to study the clustering of the dark
energy, primarily through cross-correlation with the ISW effect.  Indeed any such correlation
is a direct indication of dark energy in a spatially flat universe.
Its primary disadvantage is that
the sources are confined to a single epoch, the last scattering surface, so that
tomographic studies of the evolution of the dark energy and dark matter are impossible.  
Galaxy lensing with source redshift information 
can therefore better constrain the equation of state of the dark energy 
including potentially its evolution.

It is important to realize that Fisher parameter forecasts include statistical errors only
making the blind combination of information from disparate sources dangerous.  In particular,
the information supplied by lensing relies in large part on the accurate 
absolute calibration
of the power spectra.  On the CMB lensing side,
this involves first an accurate determination of the CMB power spectrum itself as well as
any detector or foreground power spectrum contaminants.   On the
cosmic shear side, it requires exquisite control over the myriad
systematics that enter into the measurement of shear from galaxy images.
Furthermore, Fisher forecasts are only probe the degeneracy structure locally around a fiducial model.
When error ellipses are extended in parameter space due to degeneracies, Fisher forecasts
can yield both overly optimistic or pessimistic results.  Our 
results provide the motivation for future studies that do incorporate these systematic
effects involving the combination of cosmological information from 
CMB anisotropy and gravitational lensing.

\vskip 0.5truecm
{\it Acknowledgements:}  I acknowledge useful conversations with
R. Caldwell, A.R. Cooray, D.J. Eisenstein, D. Huterer, J. Miralde-Escude, M. Zaldarriaga and the participants of
the Aspen Wide-Field Survey workshop where this work was begun.  This work was supported
by NASA NAG5-10840, DOE OJI and an Alfred P. Sloan Foundation Fellowship.

\appendix
\section{Potential Evolution and Transfer Function}

Above the sound horizon of the dark energy, defined as
\begin{equation}
s(a) = \int_z^{z_{i}} {d z' \over H(z')} c_{\rm eff} \,,
\end{equation}
where $z_i$ is some initial effectively infinite redshift,
 the Bardeen curvature
remains constant after radiation becomes negligible at some epoch 
$z_{\rm md}$.
The Newtonian curvature consequently obeys 
\begin{equation}
\Phi(k,z) = \Phi_{c}(z) \zeta(k,z_{\rm md})\,,
\end{equation}
where the decay function in the clustering regime is 
\cite{HuEis99}
\begin{equation}
\Phi_c(z) = \left( 1 - {\sqrt{\rho} \over a} \int_0^a {da \over \sqrt{\rho}}
       \right) \,,
\end{equation}
where $a=(1+z)^{-1}$.
Conversely, for scales that are much smaller than the sound horizon at
the epoch of dark energy domination
the dark energy may be considered effectively smooth for all time 
and hence the Newtonian curvature obeys 
\begin{equation}
\Phi(k,z) = \Phi_{s}(k,z) \zeta(k,z_{\rm md})\,,
\end{equation}
where 
\begin{equation}
\Phi_s''
+ 
[{5 \over 2} - {3 \over 2} w \Omega_\Lambda(z)]
\Phi_s'
+ 
{3 \over 2} [1 -w]\Omega_\Lambda(z) 
\Phi_s = 0
\end{equation}
and $'$ denotes derivatives with respect to $\ln a$. To match
solutions $\Phi_c(z_{\rm md})=\Phi_s(z_{\rm md})$ in the matter dominated epoch,
the initial conditions are set to be 
$\Phi_s(z_{\rm md})= 3/5$, $\Phi_s'(z_{\rm md})=0$.  

\begin{figure}[htb]
\centerline{\epsfxsize=3.25truein\epsffile{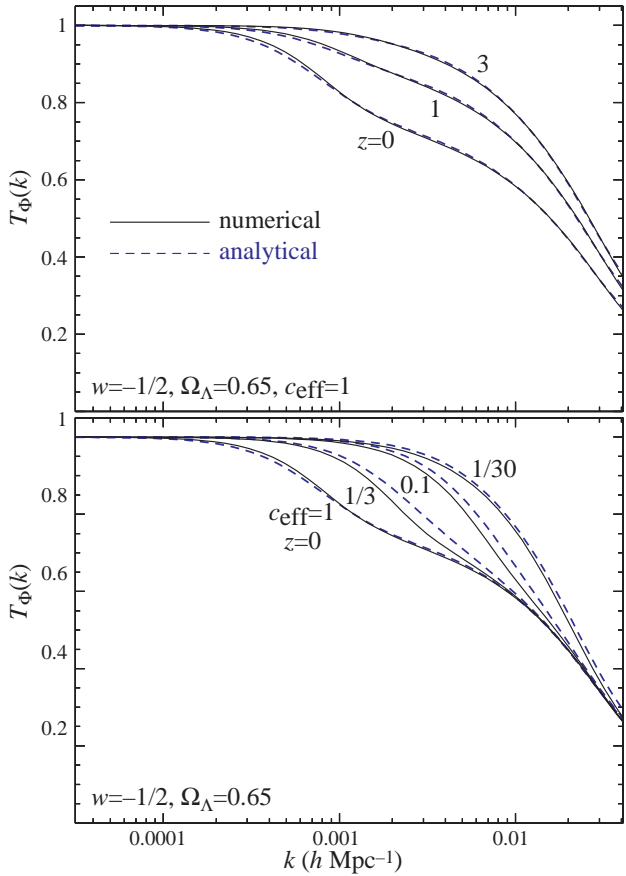}}
\vskip 0.25cm
\caption{Potential transfer function as a function of redshift
(top) and dark energy sound speed (bottom).  Solid lines represent numerical
results; dashed lines represent the analytic fits of  {\protect\cite{EisHu99}} 
supplemented by dark energy clustering. }
\label{fig:transfer}
\end{figure}

The decay function in the intermediate regime can be approximated with
a smooth interpolation of these two solutions.  First we
define the epoch of dark energy domination as
\begin{eqnarray}
{\rho_\Lambda(z_\Lambda) \over \rho_m(z_\Lambda) }
= {1 \over \pi}\,, \quad
(1+z_\Lambda)  =  \left(\pi  {\Omega_\Lambda \over \Omega_m} \right)^{-{1 
\over 3w}}\,,
\end{eqnarray}
where the solution assumes a constant equation of
state.  Next, we introduce the interpolation function
\begin{equation}
T_w(k,z) = { {1 + q^2} \over {\Phi_s/\Phi_c + q^2}}\,,
\end{equation}
where
\begin{equation}
q \equiv {k \over 2\pi} \sqrt{s(z) s(z_\Lambda)} \,.
\end{equation}
The full evolution of the potential from the matter dominated epoch 
on can be described by 
\begin{equation}
\Phi(k,z) = T_w(k,z) \Phi_s(z) \zeta(k,z_{\rm md})\,.
\end{equation}
Since in the matter dominated regime, the potential is related
to the matter density fluctuations by the Poisson equation, the 
potential transfer function asymptotically approaches a scaled
version of the matter transfer function $T_m(k)$ at high $k$
\begin{eqnarray}
T_\Phi(k,z) &=& { \Phi(k,z) \over \Phi(0,z) } {\Phi(0,z_i) \over \Phi(0,z)}\,.
\nonumber\\
	    &=& { T_w(k,z)  \over T_w(0,z) } T_m(k)\,,
\end{eqnarray}
where $T_m(k)$ is the matter transfer function assuming scale-independent
growth (a smooth dark energy component).  
Note that the true matter transfer function is still not the same
as the potential transfer function due to dark energy contributions to
the Poisson equation.  Moreover, their is no one unique matter transfer
function since it the presence of dark energy clustering the growth
of density perturbations differs between the commonly used synchronous
and comoving gauges.  The Newtonian potential transfer function is
most closely related to the comoving gauge matter transfer functions and
is in fact the density-weighted sum of the components.

\end{document}